# Toward understanding physical origin of 2175Å extinction bump in interstellar medium

Xing-Yu Ma[1], Yan-Yan Zhu[1], Qing-Bo Yan[2], Jing-Yang You[1], Gang Su[3, 1*]

[1]School of Physical Sciences, University of Chinese Academy of Sciences, Beijing 100049, China.
[2]Center of Materials Science and Optoelectronics Engineering, College of Materials Science and Optoelectronic Technology, University of Chinese Academy of Sciences, Beijing 100049, China.
[3]Kavli Institute for Theoretical Sciences, and CAS Center of Excellence in Topological Quantum Computation, University of Chinese Academy of Sciences, Beijing 100190, China

*E-mail: gsu@ucas.ac.cn (GS)

## ABSTRACT

The 2175 Å ultraviolet (UV) extinction bump in interstellar medium (ISM) of the Milky Way was discovered in 1965. After intensive exploration of more than a half century, however, its exact origin still remains a big conundrum that is being debated. Here we propose a mixture model by which the extinction bump in ISM is argued possibly relevant to the clusters of hydrogenated T-carbon (HTC) molecules ($C_{40}H_{16}$) that have intrinsically a sharp absorption peak at the wavelength 2175 Å. By linearly combining the calculated absorption spectra of HTC mixtures, graphite, $MgSiO_3$ and $Fe_2SiO_4$, we show that the UV extinction curves of optional six stars can be nicely fitted. This present work poses an alternative explanation toward understanding the physical origin of the 2175 Å extinction bump in ISM of the Milky Way.

# 1. INTRODUCTION

The interstellar extinction refers to the phenomenon that interstellar dust between earth and stars absorbs and scatters starlight. In 1965, Stecher discovered unexpectedly with sounding rocket observations that there is a 2175 Å bump feature in interstellar ultraviolet (UV) extinction curves (Stecher 1965). This extinction bump was later confirmed in tens stars by means of Orbiting Astronomical Observatory 2 (Bless & Savage 1972). Since then, several satellite observatories have helped to verify this bump feature toward many sightlines (Fitzpatrick & Massa 1986; Savage & Sembach 1996; Fitzpatrick & Massa 2007; Zafar et al. 2018). The subsequent observation in different galaxies shows that such a 2175 Å extinction bump feature in interstellar medium (ISM) is almost ubiquitous and pronounced in the Milky Way (Zafar et al. 2018; Draine 1989). It is commonly believed from observations that the 2175 Å extinction bump in ISM is from carbonaceous materials (Draine 1989). Despite intensive exploration more than a half century, however, its exact physical origin is still unclear. A few possible candidates including graphite, nongraphic carbon, diamond and polycyclic aromatic hydrocarbons were proposed, but a consensus is yet far to reach.

Earlier studies suggested that graphite grains could be responsible for the bump feature (Stecher & Donn 1965; Borg 1967; Papoular et al. 2009). Fitzpatrick and Massa discovered that the position of 2175 Å bump was rather stable, with variations in the full width at half maximum (FWHM) of the profile in reddened galaxy stars (Fitzpatrick & Massa 1986). Draine and Malhotra pointed out that different shapes and sizes of graphite may produce variations in FWHM, but these were accompanied with position changes of the UV absorption peak (Draine & Malhotra 1993), showing that graphite alone cannot explain the origin of 2175 Å extinction bump. Meanwhile, diamond was also proposed for the interstellar dust (Saslaw & Gaustad 1969). It was later argued that diamond might be the dust component to fit far ultraviolet (FUV) extinction curve and could give a smooth extinction but without distinctive features (Tielens et al. 1987), implying that diamond could not be the physical origin of the 2175 Å bump feature.

In addition, aromatic carbon or polycyclic aromatic hydrocarbons (PAHs) (Donn 1968; Vdovykin 1970) were suggested for the bump feature in ISM as well. PAHs are organic molecules composed of C and H atoms in forms of multiple aromatic rings, whose typical structures include coronene, pyrene and anthracene. It was experimentally discovered that the UV-visible absorption curve of PAH mixtures was compatible with the interstellar extinction curve, but the UV absorption spectra of anthracene and pyrene had a remarkable absorption peak at 2380 Å (Joblin, Léger & Martin 1992), which is far different from 2175 Å in the UV extinction curve. A silicate-graphite-PAH model was then proposed by adjusting the UV absorption peak of PAH using an empirical grain size distribution function to better fit the interstellar extinction curves with several parameters (Weingartner & Draine 2001). However, Mathis argued that the bump peak is sensitive to the PAH shapes (Mathis 1998), and Allamandola *et al.* reported that the absorption spectra of interstellar dust should include *$sp^3$* hybridization bonds (Allamandola et al. 1993). These studies thus pose challenges on PAHs as possible physical origin of the extinction bump, leaving active debate on this intractable issue.

In this circumstance, different candidates other than the aforesaid carbon forms to explain the extinction bump in ISM of the Milky Way should be sought for. Several years ago, a new carbon allotrope named T-carbon (Figure 1(a)) was proposed (Sheng et al. 2011), which is formed by replacing each atom in cubic diamond by a carbon tetrahedron. T-carbon possesses the same space group $Fd\bar{3}m$ as diamond and contains anisotropic $sp^3$ hybridized bonds. This novel carbon allotrope was successfully synthesized in different laboratories recently (Zhang et al. 2017; Chen et al. 2020). It is interesting to note that the measured UV-visible optical absorption spectra of T-carbon has a prominent peak at the wavelength of 2250 Å (Zhang et al. 2017), which is not far from 2175 Å. This fact strongly hints that T-carbon might be correlated with the interstellar extinction bump in some manner. As T-carbon is inclined to form under a negative pressure circumstance according to our calculations, we speculate that it is conducive to form T-carbon and/or its fragments (molecules or clusters) in interstellar space, as the latter is generally in a state of negative pressure. Because the interstellar space is full of neutral hydrogen (Kalberla et al. 2005, 2009), it is the most possible that T-carbon in interstellar space may exist in the form of molecules or clusters passivated by hydrogen atoms, forming the hydrogenated T-carbon (HTC) molecules or clusters.

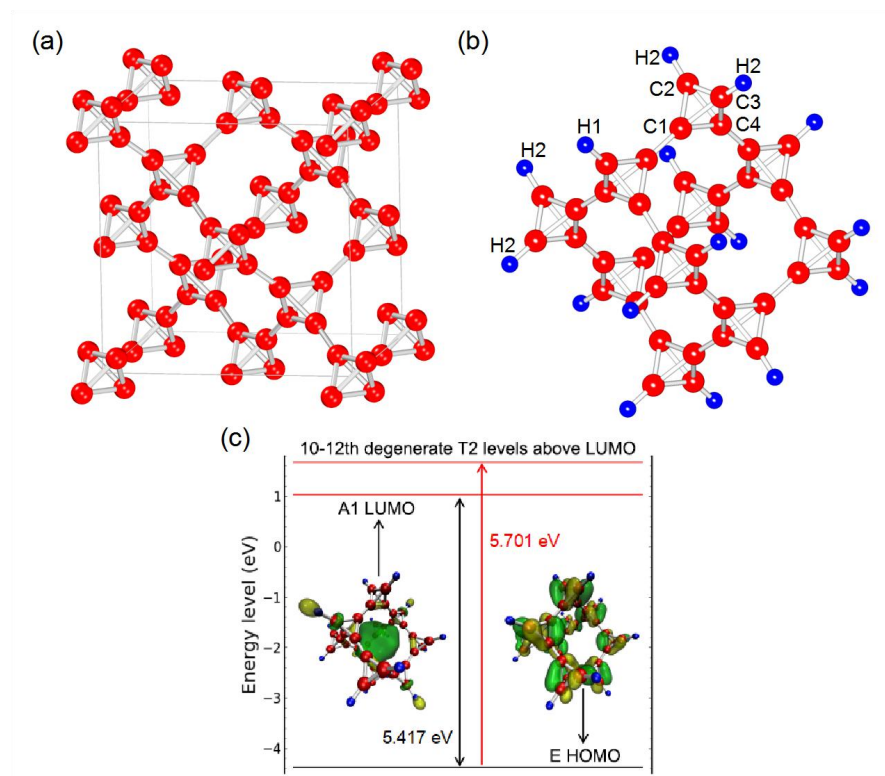

**Figure 1.** Structures of T-carbon and hydrogenated T-carbon molecule. (a) T-carbon. (b) Hydrogenated T-carbon (HTC) molecule ($C_{40}H_{16}$). The red and blue balls represent C and H atoms, respectively. **(c)** Calculated molecular orbital energy levels of HTC molecules together with the molecular orbital maps of the lowest unoccupied molecular frontier orbitals (LUMOs) and the highest occupied molecular frontier orbitals (HOMOs), with the LUMO-HOMO gap of 5.417 eV. The green and yellow regions represent the positive and negative phases of the orbital wave function. The calculations show that the 2175 Å absorption peak of HTC molecule is intrinsically generated from the excitation from the HOMO (with the orbital symmetry E) to the 10th level (degenerate in energy with the 11th and 12th levels and the orbital symmetry T2) above the LUMO (with the orbital symmetry A1).

## 2. HTC MOLECULE, GRAPHITE AND SILICATES

The structure of HTC molecule (with point group $T_d$, the same as T-carbon) is comprised of a supercell of T-carbon with passivation of hydrogen atoms, containing 40 carbon and 16 hydrogen atoms ($C_{40}H_{16}$), as presented in Figure 1(b). (The calculation methods can be found in Appendix A1.) There are three distinct bonds and four different bond angles in HTC molecules close to those in T-carbon (Table A1 in Appendix A2). The passivation of hydrogen atoms has little effect on C-C bond length and bond angles. The HTC molecule possesses $sp^3$ hybridized bonds. The molecular vibration calculations show that HTC molecules are kinetically stable (Table A2 in Appendix A4 and A5). The UV-visible absorption spectra of HTC molecules are found to have a very sharp absorption peak positioned at the wavelength of 2175 Å (Figure 2(a)), which is in accordance with the excitations from the highest occupied molecular orbitals (HOMO) to the 10th level (degenerate in energy with the 11th and 12th levels) above the lowest unoccupied molecular orbitals (LUMO) (Figure 1c), indicating that the physical origin of the interstellar UV extinction bump might be intrinsically related to HTC molecules or clusters.

We also calculated the UV absorption peaks of HTC molecule by using the PBE (Perdew et al. 1996), M06-2X (Zhao et al. 2006) and ωB97 (Chai et al. 2008) methods comparing with B3LYP and found that the main peak of the UV absorption spectra of HTC molecule by different methods is surprisingly stable and consistent (Figure A1 in Appendix A3). As the interstellar dust is usually comprised of molecules and/or clusters with various sizes, we believe that in ISM the HTC molecules in most cases may exist in the form of HTC mixtures consisting of HTC clusters with distinct sizes, leading to broadening of absorption peaks. Figure 2(a) presents the calculated sharp absorption peaks (shown by straight lines) of HTC molecule as well as those broadened absorption spectra of HTC mixtures at the absorbed wavelengths as labeled. The calculation details for HTC mixtures can refer to the next section. It is clear that the central peak of the overall broadened absorption spectra of HTC mixtures appears at 2175 Å (the red curve in Figure 2(a), which is a superposition of all broadened spectra of HTC mixtures with various sizes at all absorbed wavelengths). It should be remarked here that as the absorption peaks of HTC molecule are so small at wavelengths other than 2175 Å, they are almost covered up when superposing all broadened spectra of HTC mixtures, giving rise to a profile of the total absorption spectra of HTC mixtures centered at the wavelength 2175 Å. In addition, we also found there are eight peaks of IR spectra of HTC molecule at wavelengths 5.77, 6.23, 7.54, 8.38, 12.00, 12.65, 16.27 and 18.12 μm (The details can be found in Appendix A6) that are almost accord with the observed peaks in NGC 2023 and NGC 7023 at 5.70, 6.25, 7.63, 8.59, 11.92, 12.70, 16.42 and 17.75 μm, respectively (Sellgren et al. 2010).

Moreover, because silicate (Gilman 1969) and Mg/Fe-containing olivines (Campins & Ryan 1989; Jäger et al. 1998) as well as graphite grains might also be the components of interstellar dust, the UV absorption spectra of fayalite ($Fe_2SiO_4$), enstatite ($MgSiO_3$) and graphite should be included (Figures 2b-2d) to generate better fitting results. The calculations reveal that the UV-visible absorption spectrum of $Fe_2SiO_4$ shows a dramatic rising trend at 0-9.92 eV (0-8 μm$^{-1}$) (Figure 1(b)), which agrees with the rising linear background proposed by Fitzpatrick (Fitzpatrick 1999). The absorption spectrum of $MgSiO_3$ possesses an absorption edge at 8.31 eV (6.7 μm$^{-1}$), which shows a rising trend in the 8.31-9.92 eV (6.7-8.0 μm$^{-1}$) FUV range, being consistent with the upward trend of the FUV curve (Figure 2(c)). The UV spectra of $MgSiO_3$ and $Fe_2SiO_4$ are consistent with the previous results (Jiang &

Guo 2004; Tokár et al. 2010). The UV absorption spectrum of graphite has a distinct absorption peak of 4.30 eV (3.47 μm$^{-1}$) in the range of 0-9.92 eV (0-8 μm$^{-1}$) (Figure 2(d)).

The red lines in Figure 2 represent the smoothed spectral curves. Since the observed spectra in experiments are usually the averaged results owing to the particles or molecules randomly distributed and oriented in interstellar space, the observed data should be the averaged ones of particles or molecules of various sizes. Note that such a smoothing does not alter the position of the central peak and the shape of spectra. Thus, it is reasonable to make the smoothing when fitting to the realistic observed curves in interstellar medium. The calculation details can be found in next section and Appendix.

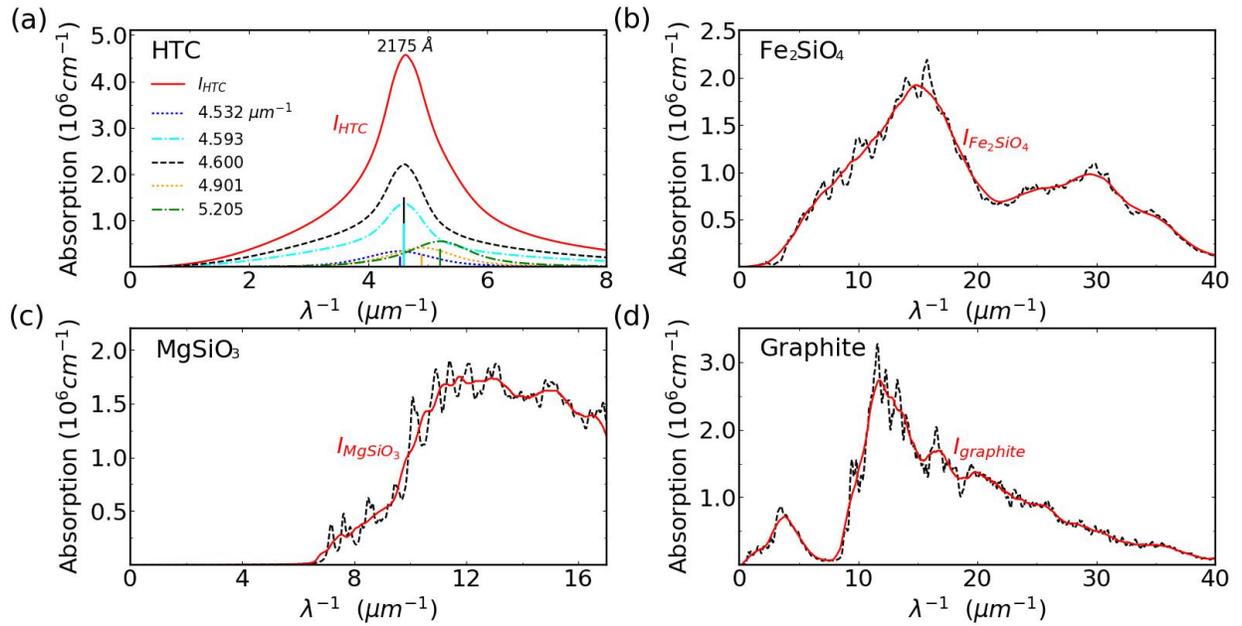

**Figure 2.** Calculated UV absorption spectra of HTC molecule, HTC mixtures, Fe$_2$SiO$_4$, MgSiO$_3$ and graphite. (a) HTC molecule, where a strikingly sharp peak appears at the wavelength of 2175 Å. The red curve represents the overall superposed broadened absorption spectra of HTC mixtures, where the peak is positioned at 2175 Å. The straight solid lines are for HTC molecule and dashed curves are for HTC mixtures at the absorbed wavelengths labeled in the legend of (a), which are obtained by Eq. (2). (b) Fe$_2$SiO$_4$, (c) MgSiO$_3$, and (d) Graphite, where red solid lines represent the smoothed spectral curves by Savitzky-Golay polynomial regression.

## 3. MODEL AND FITTING EQUATIONS

To fit the UV extinction bump curves around 3-8 μm$^{-1}$ in ISM, we take a mixture model in which four kind of primary ingredients including HTC mixtures, graphite, MgSiO$_3$ and Fe$_2$SiO$_4$ grains are included. We choose optionally six stars in the Milky Way such as BD582292, HD239689, HD239742, HD239722, HD283800 and HD147889 as examples, because they have well-documented observatory data in literature. The four stars BD582292, HD239689, HD239742 and HD239722 locate in Trumpler 37 and the extinction curves were obtained by International Ultraviolet Explorer (IUE) (Valencic et al. 2003). HD283800 is the Taurus star through the Taurus dark cloud at a distance of about 150 parsec, where its extinction curve was obtained through IUE (Whittet et al. 2004). The extinction curve of HD147899 star was found in a publication (Jenniskens & Greenberg 1993).

The fitting equation, which was frequently employed to fit 2175 Å bump in previous works (Fitzpatrick & Massa 2007; Li et al. 2001; Blasberger et al. 2017), is given by Fitzpatrick and Massa (Fitzpatrick & Massa 1988, 1990) with the absorption cross-section of Li and Draine (Li & Draine 2001) in the form of

$$I_{Fit} = \alpha_1 I_{HTC} + \alpha_2 I_{Graphite} + \alpha_3 I_{MgSiO_3} + \alpha_4 I_{Fe_2SiO_4} \qquad (1)$$

where $\alpha_1$-$\alpha_4$ are linear fitting parameters, and $I_{HTC}$, $I_{Graphite}$, $I_{MgSiO3}$ and $I_{Fe2SiO4}$ stand for the absorption intensities of HTC mixtures, graphite, MgSiO$_3$ and Fe$_2$SiO$_4$, respectively, as displayed in Figure 2.

The intensities $I_{Graphite}$, $I_{MgSiO3}$ and $I_{Fe2SiO4}$ can be obtained directly by means of first-principles calculations because graphite, MgSiO$_3$ and Fe$_2$SiO$_4$ are all assumed in solid grains, as given in Figs. 2(b)-2(d). The calculation methods can be found in Appendix A1.

To calculate $I_{HTC}$, we employ the equation of absorption intensity for mixtures given by Weingartner and Draine (Weingartner & Draine 2001) in the form of

$$I_{HTC} = (2.5\pi \lg(e)) \int \frac{dn(a)}{da} a^3 Q_{ext}(a) d\ln a \qquad (2)$$

$$dn(a) = Cn_H a^{-3.5} da \qquad (3)$$

where n(a) is the number density of HTC mixtures, a is the size of HTC mixture, $Q_{ext}$ is the extinction efficiency factor, which can be evaluated with an assumption of spherical grains by using a Mie theory derived from BHMIE (Bohren & Huffman 1983), $n_H$ is the number density of H nuclei, and C is a constant.

The extinction efficiency factor ($Q_{ext}$) can be obtained by the following equations (Bohren & Huffman 1983)

$$Q_{ext} = \frac{2}{k^2} \sum_{n=-1}^{\infty} (2n+1) \text{Re}(a_n + b_n) \qquad (4)$$

$$a_n = \frac{m\psi_n(mk)\psi_n^{'}(k) - \psi_n(k)\psi_n^{'}(mk)}{m\psi_n(mk)\xi_n^{'}(k) - \xi_n(k)\psi_n^{'}(mk)}$$

$$b_n = \frac{\psi_n(mk)\psi_n^{'}(k) - m\psi_n(k)\psi_n^{'}(mk)}{\psi_n(mk)\xi_n^{'}(k) - m\xi_n(k)\psi_n^{'}(mk)} \qquad (5)$$

$$\psi_n(x) = xj_n(x), \ \xi_n(x) = xh_n^{(1)}(x) \qquad (6)$$

$$k = 2\pi a/\lambda, \ m = \sqrt{(\sqrt{\varepsilon_{real}^2 + \varepsilon_{imag}^2} + \varepsilon_{real})/2} + i\sqrt{(\sqrt{\varepsilon_{real}^2 + \varepsilon_{imag}^2} + \varepsilon_{imag})/2} \qquad (7)$$

where k is the size parameter, $\lambda$ is the wavelength, $j_n$ is the spherical Bessel function, $h_n^{(1)}$ is the first class of spherical Hankel functions, m is the refractive indices of clusters, $\varepsilon_{real}$ and $\varepsilon_{imag}$ are the real and imaginary part of the dielectric function of mixtures, respectively.

The dielectric function of HTC mixtures can be calculated by the expression (Bohren & Huffman 1983)

$$\varepsilon = \varepsilon_{real} + i\varepsilon_{imag} = 1 + A\frac{\mu_0^2 - \mu^2}{(\mu_0^2 - \mu^2)^2 + \mu^2\gamma^2} + iA\frac{\mu\gamma}{(\mu_0^2 - \mu^2)^2 + \mu^2\gamma^2} \quad (8)$$

where $\mu$ is the photon frequency, $\mu_0$ is the photon frequency at the peak of dielectric function, A is the amplitude of the peak of dielectric function and $\gamma$ is the broadening width of the peak of dielectric functions.

The mixtures with different sizes usually have distinct parameters ($\mu_0$, A and $\gamma$) in dielectric function. To calculate these three parameters for HTC mixtures of various sizes, we suppose that the photon frequency at the peak position of dielectric function for HTC mixtures with various sizes distributes randomly near the corresponding peak position of an HTC molecule, obeying a relation

$$\mu_0(a) = \mu_p + r \times \left[\left(\frac{a - a_{molecule}}{a_{max}}\right)^\beta + \Delta\right] \quad (9)$$

where $\mu_p$ is the photon frequency at the peak of dielectric function for an HTC molecule, r is a random frequency in the range of -0.5 $\mu m^{-1}$ <r< 0.5 $\mu m^{-1}$, $a_{molecule}$ is the size of an HTC molecule (around 1 nm), $a_{max}$ is the maximum size of HTC mixtures (around 500 nm), we assume $\beta$=0.5 for BD582292, HD239689, HD239742 and HD239742, $\beta$=0.55 for HD147889, and $\Delta$ is a constant (here 0.2 is taken).

The amplitude of the peak of dielectric function of HTC mixtures is presumed to have an interpolation form

$$A(a) = \frac{A_{bulk} - A_{molecule}}{a_t - a_{molecule}} \times (a - a_{molecule}) + A_{molecule} \quad (a < a_t)$$
$$A(a) = A_{bulk} \quad (a \geq a_t) \quad (10)$$

where $A_{bulk}$ and $A_{molecule}$ are the amplitudes of the peak of dielectric function for T-carbon bulk and HTC molecule, respectively, $a_t$ is a cut-off size (200 nm) and $a_{molecule}$ is taken as above. Here we assumed that the amplitude of the peak of dielectric function is proportional to the size of system. In this way, one may see that when the size of an HTC mixture is larger than the cut-off size, the amplitude of the peak of dielectric function for HTC mixtures is close to that of T-carbon bulk. Such an assumption is compatible with the previous work (Yang et al. 2012).

The broadening width of the peak of dielectric function of HTC mixtures is also presumed to take an interpolation form

$$\gamma(a) = \frac{FWHM_{bulk} - FWHM_{molecule}}{a_t - a_{molecule}} \times (a - a_{molecule}) + FWHM_{molecule} \quad (a < a_t)$$
$$\gamma(a) = FWHM_{bulk} \quad (a \geq a_t) \quad (11)$$

where $FWHM_{bulk}$ and $FWHM_{molecule}$ represent the full width at half maximum (FWHM) of T-carbon bulk and HTC molecule, respectively. Here we adopt $FWHM_{bulk}$ = 0.4 $\mu m^{-1}$ (0.5 eV) (Zhang et al. 2017), $FWHM_{molecule}$ = 0.003 $\mu m^{-1}$ (0.00372 eV), where the broadening width of HTC molecule arises from the natural broadening. In above, we assumed that the broadening width of the peak of dielectric function is proportional to the size of system. One may note that when the size of an HTC mixture is larger than the cut-off size, the broadening width of HTC mixtures is close to that of T-carbon bulk. This assumption is also compatible with the previous work (Jasieniak, Califano & Watkins 2011).

By using the above results for HTC mixtures, we can obtain the broadened absorption spectra of HTC mixtures at the absorbed wavelengths, as given in Figure 2(a). It should be pointed out that the central peak of the overall broadened absorption spectra of HTC mixtures is still positioned at 2175 Å (namely, the red curve in Figure 2(a), which is a superposition of all broadened spectra of HTC clusters with various sizes at all absorbed wavelengths of an HTC molecule).

## 4. RESULTS AND DISCUSSION

The fitting results according to Eq. (1) are presented in Figure 3. One may see that the extinction curves of the six stars are nicely fitted by the UV absorption spectra of HTC mixtures, graphite, $MgSiO_3$ and $Fe_2SiO_4$ grains, where the adjusted coefficient of determination (goodness of fit) for fittings is about 97% with the mean squared error of about 0.016.

To show the contributing proportion of the four ingredients in fitting the UV extinction feature bump in ISM of the Milky Way, we take the column density ratio as an indicator. The column density ratio between the four ingredients can be obtained according to the formula given by Weingartner and Draine (Weingartner & Draine 2001) in the form of

$$n_{HTC} : n_{Graphite} : n_{MgSiO_3} : n_{Fe_2SiO_4} = \alpha_1 \rho_{HTC} : \alpha_2 \rho_{Graphite} : \alpha_3 \rho_{MgSiO_3} : \alpha_4 \rho_{Fe_2SiO_4} \quad (12)$$

$$\rho_{HTC} = Cn_H \int m_{molecule} \frac{a^3}{a_{molecule}^3} a^{-3.5} da \quad (13)$$

where $\alpha_1$-$\alpha_4$ are linear fitting parameters in Eq. (1), $Cn_H = 100$ cm$^{0.5}$, and $m_{molecule}$ is the mass of HTC molecule ($8.3 \times 10^{-22}$ g). The column density of HTC mixtures ($\rho_{HTC}$) can be acquired by Eq. (13), which yields 1.121 g/cm$^2$ (A detail calculation is given in Appendix A7). The column densities of $\rho_{Graphite}$=2.259 g/cm$^2$ for graphite, $\rho_{MgSiO_3}$=3.965 g/cm$^2$ for $MgSiO_3$, and $\rho_{Fe_2SiO_4}$=4.671 g/cm$^2$ for $Fe_2SiO_4$ are adopted. We choose BD582292 star as an example. The fitting results give that $\alpha_1$, $\alpha_2$, $\alpha_3$, and $\alpha_4$ are $1.14 \times 10^{-6}$, $1.46 \times 10^{-6}$, $0.72 \times 10^{-6}$ and $3.50 \times 10^{-6}$, respectively. The column density ratio of the four substances (HTC mixtures, graphite, $MgSiO_3$ and $Fe_2SiO_4$) in fitting to this star can be obtained to be 5.37%:13.87%:12.01%:68.75% by Eq. (12). It is interesting to mention that the column density ratio of carbonaceous dust (HTC mixtures and graphite) and silicate dust ($MgSiO_3$ and $Fe_2SiO_4$) in our fittings is about 0.24, which is not far from that of carbonaceous dust and silicate dust 0.38 in a previous work (Weingartner & Draine 2001). The column density ratios in other stars are listed in Table 1, which shows the results similar to BD582292 star.

The present fitting results reveal that, as shown in Figure 3, the 2175 Å UV extinction bump in ISM may be possibly contributed by HTC mixtures (~5.72% in column density ratio), graphite (~12.42%), $MgSiO_3$ (~19.83%) and $Fe_2SiO_4$ (~62.03%), while the central peak at 2175 Å in the UV extinction bump feature may be primarily attributed to HTC clusters.

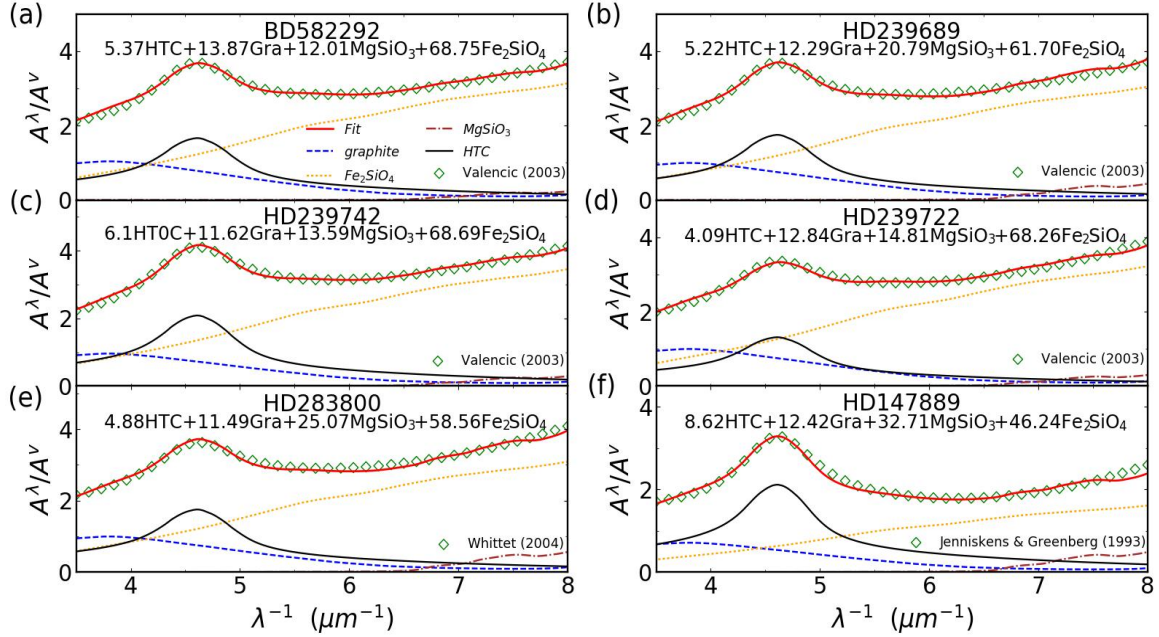

**Figure 3.** Fitting results for the normalized extinction curves of the six stars of (a) BD582292, (b) HD239689, (c) HD239742, (d) HD239742, (e) HD283800 and (f) HD147889. Diamonds denote the observed data adapted from the references as indicated. The red solid lines are fitting curves of the observed data in accordance with Eq. (1). The other lines are marked in the legend of (a). The expression in each figure shows the linear fittings with different column density ratios (in percentage) of four substances (HTC mixtures, graphite, $MgSiO_3$ and $Fe_2SiO_4$). The 2175 Å bump feature in extinction curves is nicely reproduced for the six stars.

**Table 1.** The fitting parameters, adjusted coefficients of determination, column density ratio of the four substances (HTC mixtures, graphite, $MgSiO_3$ and $Fe_2SiO_4$) in fitting the UV extinction curves of six stars BD582292, HD239680, HD239742, HD239722, HD283800 and HD147889. The corresponding mean values are included.

|  | $\alpha_1 \times 10^{-6}$ | $\alpha_2 \times 10^{-6}$ | $\alpha_3 \times 10^{-6}$ | $\alpha_4 \times 10^{-6}$ | $R^2_{adjusted}$ (%) | $n_{HTC} : n_{graphite} : n_{MgSiO_3} : n_{Fe_2SiO_4}$ (%) |
|---|---|---|---|---|---|---|
| BD582292 | 1.14 | 1.46 | 0.72 | 3.50 | 96.09 | 5.37 : 13.87 : 12.01 : 68.75 |
| HD239689 | 1.20 | 1.40 | 1.35 | 3.40 | 96.71 | 5.22 : 12.29 : 20.79 : 61.70 |
| HD239742 | 1.43 | 1.35 | 0.90 | 3.86 | 97.36 | 6.10 : 11.62 : 13.59 : 68.69 |
| HD239722 | 0.90 | 1.40 | 0.92 | 3.60 | 96.71 | 4.09 : 12.84 : 14.81 : 68.26 |
| HD283800 | 1.20 | 1.40 | 1.74 | 3.45 | 98.04 | 4.88 : 11.49 : 25.07 : 58.56 |
| HD147889 | 1.40 | 1.00 | 1.50 | 1.80 | 97.15 | 8.62 : 12.42 : 32.71 : 46.24 |
| Mean | 1.21 | 1.34 | 1.19 | 3.27 | 97.01 | 5.72 : 12.42 : 19.83 : 62.03 |

*Note.* Observation data for BD582292, HD239689, HD239742 and HD239722 are from Valencic et al. 2003. Data for HD283800 and HD147899 are from Whittet et al. 2004 and Jenniskens & Greenberg 1993, respectively.

## 5. SUMMARY AND CONCLUSIONS

In this work, we proposed a mixture model including HTC mixtures as well as graphite, MgSiO$_3$ and Fe$_2$SiO$_4$ grains to explain the 2175 Å UV extinction bump feature of the interstellar medium in the Milky Way by fitting to the extinction curves of optional six stars. Our calculations are in good agreement with the observational results, indicating that the extinction bump feature may be caused by HTC mixtures (~5.2% in column density ratio), graphite (~12.5%), MgSiO$_3$ (~19.9%) and Fe$_2$SiO$_4$ (~62.4%), which also expounds the possible physical origin of this mysterious bump feature. The present work gives an addition to possible explanations for this phenomenon, which shows that the mixtures of hydrogenated T-carbon molecules with an intrinsic sharp absorption peak at the wavelength of 2175 Å may play essential roles in explaining the UV extinction curves of interstellar dust in the Milky Way. With the present proposals we anticipate that more observations and explorations on this mixture model based on HTC clusters in ISM of the Milky Way would be conducted in future.

## ACKNOWLEDGMENTS

This work is supported in part by the National Key R&D Program of China (Grant No. 2018YFA0305800), the Strategic Priority Research Program of CAS (Grant No. XDB28000000), the NSFC (Grant No. 11834014), and Beijing Municipal Science and Technology Commission (Grant No. Z191100007219013). The calculations were performed on Era at the Supercomputing Center of Chinese Academy of Sciences.

## APPENDIX

### A1. Calculation Methods

The calculations of hydrogenated T-carbon (HTC) molecules are completed by Gaussian 09 (Frisch et al. 2016). In particular, the geometric optimization and harmonic vibrational frequencies were performed within the Becke three-parameter Lee-Yang-Parr (B3LYP) exchange-correlation functional (Becke 1993; Lee et al. 1988) and 6-311G** Gaussian basis set. Nuclear magnetic shielding tenors were calculated at GIAO-B3LYP/6-311G** level. Based on the time-dependent density-functional theory (Uchino et al. 2000; Raghavachari 2002), the ultraviolet (UV) absorption spectrum of HTC was obtained using B3LYP (13.9% exact exchange energy) and TZVP basis set. The B3LYP (13.9% exact exchange energy)/TZVP calculations also give the HOMO-LUMO gap. The calculated data are extracted by Multiwfn program (Lu & Chen 2012).

All geometric optimization of crystals was performed within the density functional theory (DFT) as implemented within the Vienna ab initio simulation package (VASP) (Kresse & Furthmuller 1996) with projected augmented wave (PAW) method (Blöchl 1994; Kresse et al. 1999) and the exchange-correlation interactions were treated within spin-polarized gradient approximation and Perdew-Burke-Ernzerhof generalized gradient approximation (PBE-GGA) (Perdew 1996). The plane-wave cutoff energy is taken as 500 eV and 8×8×8 *k*-points sampling are used for the ground state calculation. The lattice geometries and atomic positions were fully relaxed until the energy and force were converged to 10$^{-6}$ eV and 0.01 eV/Å.

In practical calculations, the on-site Coulomb correlated potential for Fe ions was chosen as U=4.5 eV, as derived for fayalite (Cococcioni et al. 2003), and Hund's exchange J=0.9 eV was obtained using the Racah parameters for Fe ions (Zaanen 1990).

The GW approximation (Hedin 1965, 1999) was applied to obtain the quasiparticle (QP) energies via the perturbative solution to the Dyson equation

$$[-\frac{\hbar^2}{2m_e}\nabla^2 + V_{ion} + V_H + \sum E_{nk}^{QP}]\psi_{nk}^{QP} = E_{nk}^{QP}\psi_{nk}^{QP} \tag{14}$$

where $m_e$ is the mass of electron, $\hbar$ is the reduced Planck constant, $V_{ion}$ is the electrostatic potential contributed by ions, $V_H$ is the Hartree potential, and $E_{nk}^{QP}$ is the quasi-particle energy.

For MgSiO$_3$ crystal, the optical excitation energies and exciton wave functions are determined through Bethe-Salpeter equation (BSE) (Salpeter & Bethe 1951; Strinati 1984; Rohlfing & Louie 2000)

$$(E_{ck}^{QP} - E_{vk}^{QP})A_{vck}^S + \sum_{v'c'k'}\langle vck|K^{eh}|v'c'k'\rangle A_{v'c'k'}^S = \Omega^S A_{vck}^S \tag{15}$$

where $E_{ck}^{QP}$ and $E_{vk}^{QP}$ are the quasi-particle energies for conduction and valence bands, $K^{eh}$ and $\Omega^S$ are the electron-hole interaction kernel and excitation energy, respectively.

The imaginary part of MgSiO$_3$ crystal is evaluated from the excitation energies and exciton wave functions with the following expression (Rohlfing & Louie 2000)

$$\varepsilon(\omega) = \frac{16\pi e^2}{\omega^2}\sum_S |\vec{\lambda}\langle 0|\vec{v}|S\rangle|^2 \delta(\omega - \Omega^S) \tag{16}$$

where $\vec{\lambda}$ and $\vec{v}$ is the polarization vector of the incident light and velocity operator, respectively, and $\langle 0|\vec{v}|S\rangle$ is the transition matrix element.

The imaginary part of Fe$_2$SiO$_4$ and graphite are given by

$$\varepsilon_{\mu v}(\omega) = \frac{4e^2\pi^2}{V}\lim_{q\to 0}\frac{1}{q^2}\sum_{c,v,k} 2w_k\delta(\varepsilon_{ck} - \varepsilon_{vk} - \omega) \times \langle \mu_{ck+e_\mu q}|\mu_{vk}\rangle\langle \mu_{ck+e_\mu q}|\mu_{vk}\rangle^* \tag{17}$$

where c and v denote the conduction and valence states, respectively, V is the cell volume, $e_\mu$ are unit vectors for three Cartesian directions, $w_k$ are k-points weight, and $\mu_{ck}$ is the cell periodic part of the wave function at the point k.

The absorption coefficient $\alpha(\omega)$ is calculated as follows (Saha et al. 2000)

$$\alpha(\omega) = \sqrt{2}\omega\left[\sqrt{\varepsilon_1^2(\omega) + \varepsilon_2^2(\omega)} - \varepsilon_1(\omega)\right]^{1/2} \tag{18}$$

where $\varepsilon_1(\omega)$ and $\varepsilon_2(\omega)$ are the real and imaginary parts of the dielectric function, respectively.

The kinetic energy cutoff for GW and BSE is 400 eV, 10 occupied and 10 unoccupied orbitals are used to build the electron-hole interaction kernel.

The adjusted coefficient of determination ($R_{adjusted}^2$), employed to evaluate the goodness of fit, is defined as

$$R^2 = 1 - \frac{\sum_i (y_i^{Exp} - y_i^{Fit})^2}{\sum_i (y_i^{Exp} - \overline{y_i^{Exp}})^2} \tag{19}$$

$$R_{adjusted}^2 = 1 - \frac{(1-R^2)(N-1)}{N-1-p} \tag{20}$$

where $y_i^{Exp}$ is the experimental observation value, $y_i^{Fit}$ is the fitting value, N is the number of data and p is the number of fitting parameters. The closer to 1 the value of $R_{adjusted}^2$, the better the fitting degree on experimental values.

The mean squared error (MSE) represents the expected value of the square of the difference between the fitting values and experimental data, which is defined as

$$\text{MSE} = \frac{1}{N}\Sigma_i^N (y_i^{Exp} - y_i^{Fit})^2 \tag{21}$$

## A2. Structural parameters of HTC molecule

In Figure 1(b) of the text, the chemical environments of C1 and C4 atoms in each carbon tetrahedron, C2 and C3 atoms, and H1 and H2 atoms in an HTC molecule are the same (Figure 1(b)), respectively. Electrons of LUMOs and HOMOs (Figure 1(c)) are mainly concentrated on the interior of the HTC molecule and C-C bonds in tetrahedrons, respectively. Table A1 gives the structural parameters of an HTC molecule, including the bond lengths of C-C and C-H atoms, and the bond angles between C atoms at different positions.

**Table A1.** The structural parameters of HTC molecule

|  | Bond length (Å) |
|---|---|
| C-H1 | 1.702 |
| C-H2 | 1.701 |
| C1-C2 | 1.485 |
| C1-C4 | 1.490 |
| C2-C3 | 1.482 |
| C4-C5 | 1.424 |
|  | Bond angle (°) |
| C1-C2-C3 | 60.07 |
| C1-C3-C2 | 60.20 |
| C1-C3-C4 | 59.90 |
| C2-C1-C3 | 59.87 |
| C3-C4-C5 | 144.68 |

## A3. Comparison between different DFT methods

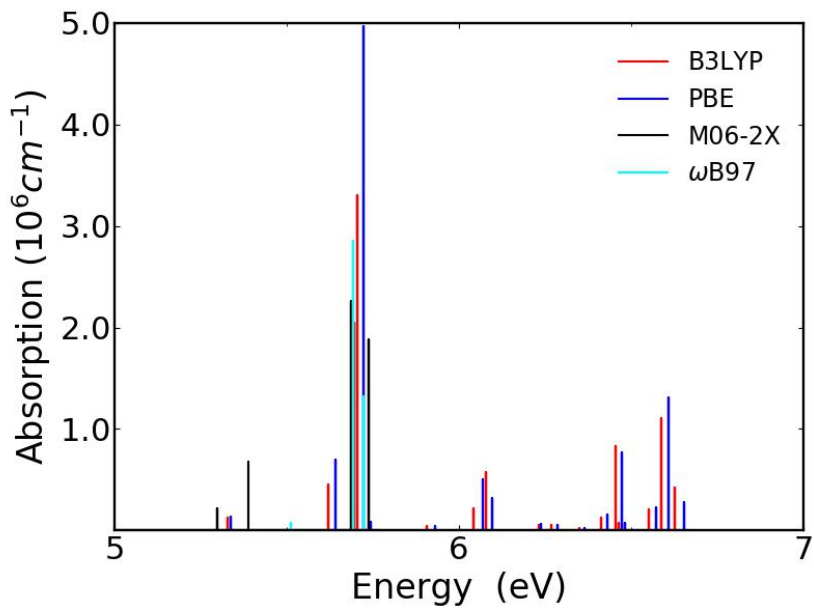

**Figure A1.** Calculated UV absorption spectra of HTC molecule by different methods B3LYP, PBE, M06-2X and ωB97.

We calculated the UV absorption spectra of HTC molecule by using the PBE (Perdew et al. 1996), M06-2X (Zhao et al. 2006) and ωB97 (Chai et al. 2008) methods comparing with B3LYP, and found that the main peak of the UV absorption spectra of HTC molecule by different methods is surprisingly stable (PBE: 216.6 nm, M06-2X: 218.0 nm, ωB97: 217.9 nm and B3LYP: 217.5 nm), as presented in Figure A1. Therefore, these results show that our calculations are very reliable.

# A4. Excited States, molecule orbital transition, orbital symmetry of main transition and contributions of transition for HTC molecule

In Figure 2(a), we have listed five main peaks with the intensity and corresponding excitation energies, oscillator strengths, excited states, molecular orbitals transition and contributions of transition. As shown in Figure 2(a) and Table A2, the 2175 Å absorption peak corresponds to the excitation energy 5.6956 and 5.7052 eV, where the latter has the greatest contribution, and the three excited states 27, 28 and 29th are the origin of excitation energy 5.7052 eV that is mainly generated from the excitation from HOMO (127, 128th) (with orbital symmetry E) to the 10th level (degenerate in energy with the 11th and 12th levels with orbital symmetry T2) above LUMO (with orbital symmetry A1).

**Table A2.** Excitation energy, oscillator strengths, excited states, molecular orbitals transition, orbital symmetry of main transition and contributions of transition for an HTC molecule

| Excitation energy(eV) | Oscillator strengths | Excited states | Molecular orbital transition and orbital symmetry of main transition | Contribution (%) |
|---|---|---|---|---|
| 5.6199 | 0.0225 | 18 | 126 -> 129 | 12.2473 |
| | | | 127 -> 133 | 33.6725 |
| | | | 127 -> 140 | 11.1685 |
| | | | 128 -> 133 | 15.4846 |
| | | | 128 -> 140 | 24.2877 |
| | | 19 | 124 -> 129 | 12.1584 |
| | | | 127 -> 141 | 35.0167 |
| | | | 128 -> 134 | 48.5487 |
| | | 20 | 125 -> 129 | 12.1712 |
| | | | 127 -> 135 | 39.5605 |
| | | | 127 -> 139 | 6.7029 |
| | | | 128 -> 135 | 9.2924 |
| | | | 128 -> 139 | 28.5345 |
| 5.6956 | 0.1018 | 24 | **124(T2) -> 129(A1)** | **60.3022** |
| | | | 125 -> 129 | 3.9649 |
| | | | 126 -> 129 | 13.8664 |
| | | | 127 -> 133 | 2.0779 |
| | | | 128 -> 134 | 13.1226 |
| | | 25 | 124 -> 129 | 11.6905 |
| | | | 125 -> 129 | 2.5910 |
| | | | **126(T2) -> 129(A1)** | **63.8518** |
| | | | 127 -> 133 | 9.5677 |
| | | | 128 -> 133 | 4.3998 |
| | | 26 | 128 -> 134 | 2.5439 |
| | | | 124 -> 129 | 6.1411 |
| | | | **125(T2) -> 129(A1)** | **71.5782** |
| | | | 127 -> 135 | 12.6796 |
| | | | 128 -> 135 | 2.9783 |
| | | | 126 -> 129 | 5.9430 |
| | | | 127 -> 133 | 19.5163 |
| | | 27 | 127 -> 137 | 2.4589 |
| | | | 127 -> 140 | 17.9592 |
| | | | 128 -> 133 | 8.9752 |
| | | | **128(E) -> 140(T2)** | **39.0534** |
| **5.7052** | **0.1643** | 28 | 124 -> 129 | 5.8256 |
| | | | **127(E) -> 141(T2)** | **55.5964** |
| | | | 128 -> 134 | 27.7855 |
| | | | 128 -> 138 | 3.5007 |
| | | 29 | 125 -> 129 | 5.8011 |
| | | | 127 -> 135 | 22.5228 |
| | | | 127 -> 136 | 2.8379 |
| | | | 127 -> 139 | 10.5864 |
| | | | 128 -> 135 | 5.2904 |
| | | | **128(E) -> 139(T2)** | **45.0680** |
| | | 53 | 125 -> 132 | 36.5222 |
| | | | 126 -> 131 | 36.5222 |
| | | | 127 -> 149 | 17.5208 |
| 6.0773 | 0.0289 | 54 | 124 -> 132 | 36.5222 |
| | | | 126 -> 130 | 36.5205 |
| | | | 127 -> 151 | 3.3504 |
| | | | 128 -> 151 | 14.2631 |
| | | 55 | 124 -> 131 | 36.5342 |
| | | | 125 -> 130 | 36.5342 |
| | | | 127 -> 150 | 5.5504 |
| | | | 128 -> 150 | 12.0688 |
| | | 115 | 121 -> 135 | 4.4182 |
| | | | 121 -> 136 | 7.9162 |
| | | | 122 -> 134 | 4.4182 |
| | | | 122 -> 138 | 7.9162 |
| | | | 124 -> 136 | 14.9506 |
| | | | 124 -> 139 | 2.283 |
| | | | 125 -> 138 | 14.9506 |
| | | | 125 -> 141 | 2.283 |
| | | | 127 -> 153 | 9.7003 |
| | | | 127 -> 156 | 5.8332 |
| | | | 128 -> 153 | 4.4611 |
| | | | 128 -> 156 | 12.6857 |
| 6.4547 | 0.0414 | 116 | 122 -> 133 | 4.3838 |
| | | | 122 -> 137 | 7.8543 |
| | | | 123 -> 135 | 4.3838 |
| | | | 123 -> 136 | 7.8543 |
| | | | 125 -> 137 | 14.8338 |
| | | | 125 -> 140 | 2.2655 |
| | | | 126 -> 136 | 14.8338 |
| | | | 126 -> 139 | 2.2655 |
| | | | 127 -> 157 | 18.2795 |
| | | | 128 -> 154 | 13.9783 |
| | | 117 | 121 -> 133 | 4.3832 |
| | | | 121 -> 137 | 7.8535 |
| | | | 123 -> 134 | 4.3832 |
| | | | 123 -> 138 | 7.8535 |
| | | | 124 -> 137 | 14.8316 |
| | | | 124 -> 140 | 2.265 |
| | | | 126 -> 138 | 14.8316 |

| | | |
|---|---|---|
| 126 -> 141 | | 2.265 |
| 127 -> 155 | | 11.3774 |
| 127 -> 158 | | 3.4948 |
| 128 -> 155 | | 2.6722 |
| 128 -> 158 | | 14.8774 |

*Note.* The 127th and 128th molecular orbitals are the highest occupied molecular frontier orbitals (HOMOs) that are degenerate in energy. The 129th molecular orbital is the lowest unoccupied molecular frontier orbital (LUMO). The greatest contribution for molecular orbital transition and orbital symmetry are shown in boldface font.

## A5. Vibrational frequencies of HTC molecule

Table A3 presents the vibrational frequencies of the HTC molecule, showing that it is kinetically stable, as no imaginary mode is observed.

**Table A3.** The vibrational frequencies of the HTC molecule

| Mode | Frequencies (cm$^{-1}$) | Mode | Frequencies (cm$^{-1}$) |
| --- | --- | --- | --- |
| 1 | 50.2855 | 82 | 814.6955 |
| 2 | 50.2855 | 83 | 818.8367 |
| 3 | 50.2855 | 84 | 820.0789 |
| 4 | 63.5687 | 85 | 820.0789 |
| 5 | 63.5687 | 86 | 820.0789 |
| 6 | 71.1983 | 87 | 834.9854 |
| 7 | 71.1983 | 88 | 834.9854 |
| 8 | 71.1983 | 89 | 837.9844 |
| 9 | 115.9677 | 90 | 837.9844 |
| 10 | 115.9677 | 91 | 837.9844 |
| 11 | 115.9677 | 92 | 843.5936 |
| 12 | 184.9283 | 93 | 843.5936 |
| 13 | 184.9283 | 94 | 843.5936 |
| 14 | 184.9283 | 95 | 848.7936 |
| 15 | 197.4205 | 96 | 848.7936 |
| 16 | 218.7256 | 97 | 850.3619 |
| 17 | 218.7256 | 98 | 850.3619 |
| 18 | 218.7256 | 99 | 850.3619 |
| 19 | 270.2911 | 100 | 852.0032 |
| 20 | 299.6154 | 101 | 859.5882 |
| 21 | 299.6154 | 102 | 859.5882 |
| 22 | 299.6154 | 103 | 859.5882 |
| 23 | 314.4171 | 104 | 860.0699 |
| 24 | 314.4171 | 105 | 860.0699 |
| 25 | 317.1261 | 106 | 860.0699 |
| 26 | 317.1261 | 107 | 1113.847 |
| 27 | 317.1261 | 108 | 1113.847 |
| 28 | 329.7647 | 109 | 1113.847 |
| 29 | 329.7647 | 110 | 1116.0373 |
| 30 | 329.7647 | 111 | 1116.0373 |
| 31 | 332.2067 | 112 | 1116.0373 |
| 32 | 332.2068 | 113 | 1116.1246 |
| 33 | 339.6876 | 114 | 1116.1246 |
| 34 | 339.6876 | 115 | 1116.1246 |
| 35 | 339.6876 | 116 | 1120.8069 |
| 36 | 345.3286 | 117 | 1140.9626 |
| 37 | 543.431 | 118 | 1140.9626 |
| 38 | 543.431 | 119 | 1145.0886 |
| 39 | 543.431 | 120 | 1145.0886 |
| 40 | 553.6009 | 121 | 1145.0886 |
| 41 | 553.6009 | 122 | 1163.3411 |
| 42 | 553.6009 | 123 | 1163.3411 |
| 43 | 567.6182 | 124 | 1163.3411 |
| 44 | 567.6182 | 125 | 1179.8414 |
| 45 | 569.0809 | 126 | 1228.2396 |
| 46 | 569.0809 | 127 | 1228.2396 |
| 47 | 569.0809 | 128 | 1228.2396 |
| 48 | 573.9239 | 129 | 1297.0191 |
| 49 | 588.8335 | 130 | 1297.0191 |
| 50 | 588.8335 | 131 | 1361.7527 |
| 51 | 588.8335 | 132 | 1361.7527 |
| 52 | 590.3399 | 133 | 1361.7527 |
| 53 | 590.3399 | 134 | 1474.237 |
| 54 | 590.3399 | 135 | 1548.9786 |
| 55 | 622.0563 | 136 | 1548.9786 |
| 56 | 622.0563 | 137 | 1548.9786 |
| 57 | 633.7698 | 138 | 1651.6013 |
| 58 | 633.7698 | 139 | 1651.6013 |
| 59 | 633.7698 | 140 | 1651.6013 |
| 60 | 635.2303 | 141 | 1701.7048 |
| 61 | 671.6014 | 142 | 1701.7048 |
| 62 | 671.6014 | 143 | 1776.1358 |
| 63 | 671.6014 | 144 | 1776.1358 |
| 64 | 675.7037 | 145 | 1776.1358 |
| 65 | 675.7037 | 146 | 1804.9926 |
| 66 | 696.5178 | 147 | 3316.7647 |
| 67 | 704.7426 | 148 | 3316.7647 |
| 68 | 704.7426 | 149 | 3316.7647 |
| 69 | 704.7426 | 150 | 3316.8296 |
| 70 | 720.6237 | 151 | 3316.8296 |
| 71 | 720.6237 | 152 | 3316.8296 |
| 72 | 720.6237 | 153 | 3321.1971 |
| 73 | 723.899 | 154 | 3321.1971 |
| 74 | 723.899 | 155 | 3321.1971 |
| 75 | 811.2774 | 156 | 3321.2374 |
| 76 | 811.2774 | 157 | 3336.834 |
| 77 | 813.1699 | 158 | 3336.834 |
| 78 | 813.1699 | 159 | 3336.8846 |
| 79 | 813.1699 | 160 | 3336.8846 |
| 80 | 814.6955 | 161 | 3336.8846 |
| 81 | 814.6955 | 162 | 3337.0626 |

## A6. Infrared, Raman and NMR spectra of the HTC molecule

The all calculations are performed by Gaussian 09 (Frisch et al. 2016). In particular, the infrared (IR) and Raman spectra were calculated within the Becke three-parameter Lee-Yang-Parr (B3LYP) exchange-correlation functional (Becke 1993; Lee et al. 1988) and 6-311G** Gaussian basis set. Nuclear magnetic resonance (NMR) spectra was calculated at GIAO-B3LYP/6-311G** level.

Figure A2 presents the infrared (IR), Raman and NMR spectra of an HTC molecule. The IR and Raman spectra spread up to more than 3 μm. The region of a sharp peak at 3.1 μm corresponds to the C-H stretching modes as the same mode in Raman spectra, and the region more than 5 μm corresponds to the C-C stretching, C-C-C bending, and C-C-H bending modes in which active modes are different for IR and Raman spectra, as shown in Figs. A2 (a) and (b).

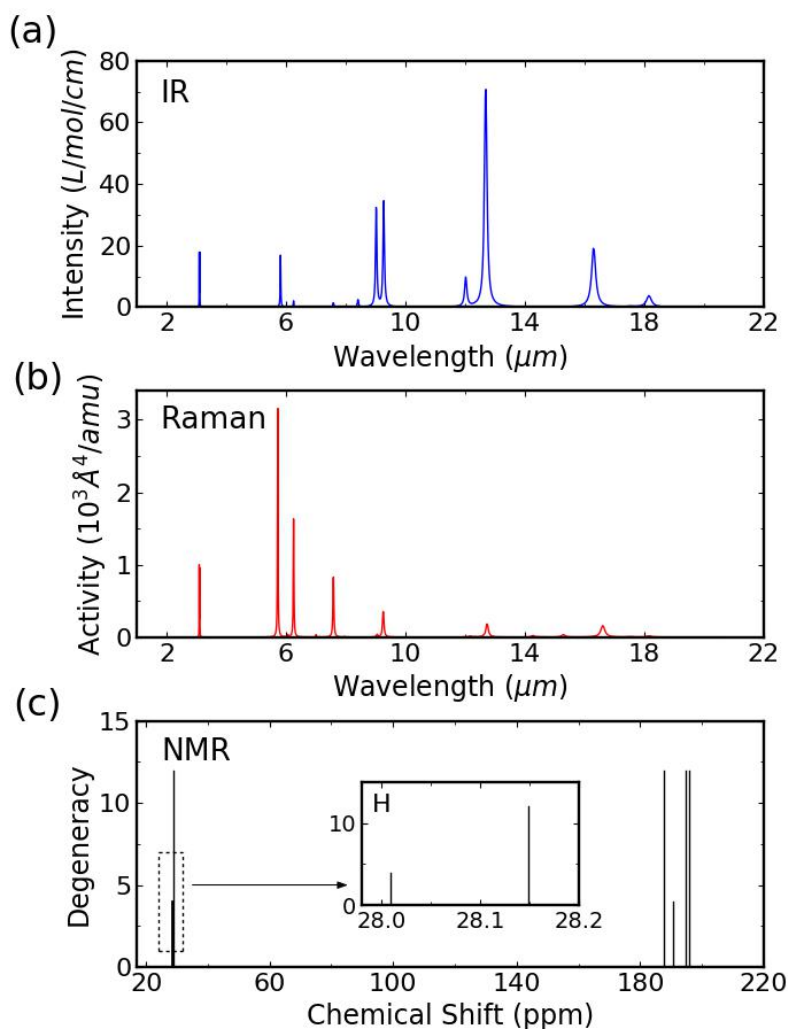

**Figure A2.** The infrared (IR), Raman and NMR spectra of HTC molecule. (a) IR, which is scaled by a factor of 0.9679. (b) Raman. (c) NMR, where the inset shows the chemical shifts of H atoms.

Figure A2 (c) shows the NMR spectra of the HTC molecule. There are also two regions. Two peaks appear in the left region whose relative degeneracy is 1:3, which could be ascribed to two unique types of H atoms [total 16 H atoms (1×4, 1×12)]. The lower peak is ascribed to H1 atoms, and another peak is ascribed to H2 atoms. Four peaks appear in the right region, three of which have the same degeneracy, and the other has a third of the degeneracy, which is ascribed to the four unique types of C atoms [40 $sp^3$ C atoms (3×12, 1×4)], where a×b represents "a" unique types, and in one type, there are "b" symmetrical equivalent atoms.

As shown in Fig. A2 (a), we found eight peaks of IR spectra of HTC molecule at wavelength 5.77, 6.23, 7.54, 8.38, 12.00, 12.65, 16.27 and 18.12 μm that are almost consistent with the observed peaks in NGC 2023 and NGC 7023 at 5.70, 6.25, 7.63, 8.59, 11.92, 12.70, 16.42 and 17.75 μm, respectively (Sellgren et al. 2010). The other two peaks of HTC molecule at 9.00 and 9.27 μm may be covered by the IR spectra of other stuffs (e.g. PAH molecules), which is like the case for the peak at 7.04 μm of C60 molecule (Sellgren et al. 2010).

## A7. Calculation of column density of HTC mixtures

The column density of HTC mixtures ($\rho_{HTC}$) can be calculated according to the formula given by Weingartner and Draine (Weingartner & Draine 2001)

$$\rho_{HTC} = Cn_H \int m_{molecule} \frac{a^3}{a_{molecule}^3} a^{-3.5} da \tag{22}$$

$$= Cn_H \times \frac{m_{molecule}}{a_{molecule}^3} \times 2(\sqrt{a_{max}} - \sqrt{a_{molecule}}) \tag{23}$$

where we take $Cn_H = 100$ cm$^{0.5}$, $m_{molecule}$ is the mass of HTC molecule (8.3×10$^{-22}$ g), $a_{molecule}$ is the size of HTC molecule (1 nm), and $a_{max}$ is the maximum size of HTC mixtures (around 500 nm). By Eq. (14), we can obtain the column density of HTC mixtures being about 1.121 g/cm$^2$. If we take $Cn_H = 3.23\times10^{-4}$ cm$^{0.5}$, $\rho_{HTC}$ will be 3.6×10$^{-6}$ g/cm$^2$. However, the fitting results give that $\alpha_1$, $\alpha_2$, $\alpha_3$, and $\alpha_4$ are 0.3426, 1.46×10$^{-6}$, 0.72×10$^{-6}$ and 3.50×10$^{-6}$, respectively. The column density ratio of the four substances (HTC mixtures, graphite, MgSiO$_3$ and Fe$_2$SiO$_4$) in fitting to this star can be obtained to be 5.37%:13.87%:12.01%:68.75% by Eq. (12), which is consistent with the result of $Cn_H = 100$ cm$^{0.5}$. Thus, the value of $Cn_H$ does not affect the column density ratio.

## DATA AVAILABILITY

The data underlying this article will be shared on reasonable request to the corresponding author.